\documentclass[runningheads]{llncs}
\usepackage{url}
\usepackage{paralist}
\usepackage{booktabs}
\usepackage{tabularx}
\usepackage{vcell}
\usepackage{multirow}
\usepackage{colortbl}
\usepackage{xcolor}
\usepackage{graphicx}
\usepackage{enumitem}
\usepackage{amsmath}
\usepackage{mdframed}
\usepackage{hyperref}
\usepackage[english]{babel}
\usepackage[autostyle=true]{csquotes}
\usepackage{tcolorbox}
\usepackage{tikz}

\usepackage{float}

\newcommand{\autour}[1]{\tikz[baseline=(X.base)]\node [draw=black,fill=cyan!20,thick,rectangle,rounded corners=4pt,text depth=0pt] (X) {#1};}
   
\usepackage{array}
\usepackage{makecell}

\renewcommand{\figurename}{Fig.}
\renewcommand{\tablename}{Tab.}

\newcommand{\blue}[1]{\textbf{\textcolor{blue}{#1}}}

\newcommand{\orange}[1]{\textbf{\textcolor{orange}{#1}}}

\usepackage{butterma}

\begin{document}
\title{Automatic Detection of Causality in Requirement Artifacts: the CiRA Approach}
\titlerunning{CiRA: Causality Detection in Requirement Artifacts}
\authorrunning{J. Fischbach et al.}

\author{Jannik Fischbach\inst{1} \and
Julian Frattini\inst{2} \and
Arjen Spaans\inst{1} \and
Maximilian Kummeth\inst{1} \and
Andreas Vogelsang\inst{3} \and Daniel Mendez\inst{2,4} \and Michael Unterkalmsteiner\inst{2}
}
%
%
\institute{Qualicen GmbH, Germany, \email{\{firstname.lastname\}@qualicen.de} \and
Blekinge Institute of Technology, Sweden, \email{\{firstname.lastname\}@bth.se} \and
University of Cologne, Germany, \email{vogelsang@cs.uni-koeln.de} \and
fortiss GmbH, Germany, \email{mendez@fortiss.org}}
\maketitle
\thispagestyle{electronic}             
\begin{abstract}
[\textbf{Context \& motivation:}]
System behavior is often expressed by causal relations in requirements (e.g., \textit{If event 1, then event 2}). Automatically extracting this embedded causal knowledge supports not only reasoning about requirements dependencies, but also various automated engineering tasks such as seamless derivation of test cases. However, causality extraction from natural language (NL) is still an open research challenge as existing approaches fail to extract causality with reasonable performance. [\textbf{Question/problem:}]
We understand causality extraction from requirements as a two-step problem: First, we need to detect if requirements have causal properties or not. Second, we need to understand and extract their causal relations. At present, though, we lack knowledge about the form and complexity of causality in requirements, which is necessary to develop a suitable approach addressing these two problems. [\textbf{Principal ideas/results:}]
We conduct an exploratory case study with 14,983 sentences from 53 requirements documents originating from 18 different domains and shed light on the form and complexity of causality in requirements. Based on our findings, we develop a tool-supported approach for causality detection (CiRA, standing for Causality in Requirement Artifacts). This constitutes a first step towards causality extraction from NL requirements.  [\textbf{Contribution:}]
We report on a case study and the resulting tool-supported approach for causality detection in requirements. Our case study corroborates, among other things, that causality is, in fact, a widely used linguistic pattern to describe system behavior, as about a third of the analyzed sentences are causal. We further demonstrate that our tool CiRA achieves a macro-F\textsubscript{1} score of 82~\% on real word data and that it outperforms related approaches with an average gain of 11.06~\% in macro-Recall and 11.43~\% in macro-Precision. Finally, we disclose our open data sets as well as our tool to foster the discourse on the automatic detection of causality in the RE community.

\keywords{Causality \and Case Study \and Requirements Engineering \and Natural Language Processing}
\end{abstract}

\section{Introduction}
\vspace{-.2cm} 
System behavior is usually described by causal relations, e.g. \enquote{A confirmation message shall be shown if the system has successfully processed the data.} Hence, causal relations are often inherently embedded in the textual descriptions of requirements. Understanding and extracting these causal relations offers great potential for Requirements Engineering (RE); for instance, by supporting the automated derivation of test cases and by facilitating reasoning about dependencies between requirements~\cite{fischbach2020}. However, automated causality extraction from requirements is still challenging for two reasons. First, requirements are mostly expressed by unrestricted natural language (NL) so that the system behavior is specified in arbitrarily complex ways. Second, causality can occur in different forms~\cite{blanco08} such as \textit{marked}/\textit{unmarked} or \textit{explicit}/\textit{implicit} which makes it difficult to identify and extract the causes and effects. Existing approaches~\cite{Asghar16} fail to extract causality from NL with a performance that allows for use in practice. Therefore, we argue for the need of a novel method for the extraction of causality from requirements. We understand causality extraction as a two-step problem: We first need to detect whether requirements contain causal relations. Second, if they contain causal relations, we need to understand and extract them. To address both problems, we have to comprehend in which form and complexity requirements causality occurs in practice. This enables us to develop efficient approaches for the automated identification and extraction of causal relations. However, empirical evidence on causality in requirements is presently still weak. In this paper, we report on how we addressed this research gap and make the following contributions (C):

\begin{compactitem}
  \item \textbf{C 1}: We report on an exploratory case study where we analyze form and complexity of causality in requirements based on 14,983~sentences emerging from 53~requirement documents. These documents originate from 18~different domains. We corroborate, for example, that causality tends to occur, in fact, in \textit{explicit} and \textit{marked} form, and that about 28~\% of the analyzed sentences contain causal knowledge about the expected system behavior. This strengthens our confidence in the relevance of our approach.
  \item \textbf{C 2}: We present our tool-supported approach named CiRA (\textbf{C}ausality detection \textbf{i}n \textbf{R}equirement \textbf{A}rtifacts), which forms a first step towards causality extraction from NL requirements. We train and empirically evaluate CiRA using the pre-analyzed data set and achieve an macro-F\textsubscript{1} score of 82~\%. Compared to baseline systems that classify causality based on the presence of certain cue phrases, or shallow ML models, CiRA leads to an average performance gain of 11.43~\% in macro-Precision and 11.06~\% in macro-Recall. 
  \item \textbf{C 3}: To strengthen transparency and facilitate replication, we disclose our tool, code, and data set used in the case study.\footnote[1]{A demo of CiRA can be accessed at \url{cira.diptsrv003.bth.se}. Our code and annotated data sets can be found at \url{https://github.com/fischJan/CiRA}.}
\end{compactitem}
\vspace{-.2cm}
\section{Terminology}\label{terminology}
\vspace{-.1cm} 
Causality represents a semantic relation that has been studied by various disciplines, e.g. by psychology~\cite{psycho03}. Before we can investigate in which form causality occurs in requirements, we must first understand what causality actually means.
\paragraph{Concept of Causality} 
Causality is a relation between two events: a causing event (the \emph{cause}) and a caused event (the \emph{effect}). An event is \enquote{any situation (including a process or state) that happens or occurs either instantaneously (punctual) or during a period of time (durative)}~\cite{mostafazadeh16}. The connection between causes and effects is counterfactual~\cite{Lewis1973a}: If a cause $c_1$ did not occur, then an effect $e_1$ could not have occurred either. Consequently, a causal relation requires that the effect may only occur \emph{if and only if} the cause has occurred. Therefore, in the view of Boolean algebra, a causal relation can be interpreted as an equivalence between a cause and effect ($c_1 \iff e_1$). If the cause is true, the effect is true and if the cause is false, the effect is also false. The relation between a cause and effect can be defined in three different ways~\cite{Wolff07}: as a \textit{cause}, \textit{enable} or \textit{prevent} relationship.

\begin{compactitem}
    \item \textbf{$c_1$ causes $e_1$}: If $c_1$ occurs, $e_1$ also occurs ($c_1 \iff e_1$). This can be illustrated by REQ 1: \enquote{After the user enters a wrong password, a warning window shall be shown.} In this case, the wrong input is the trigger to display the window.
    \item \textbf{$c_1$ enables $e_1$}:  If $c_1$ does not occur, $e_1$ does not occur either ($e_1$ is not enabled). REQ 2: \enquote{As long as you are a student, you are allowed to use the sport facilities of the university ($c_1 \iff e_1$).} Only the student status enables to do sports on campus. 
    \item \textbf{$c_1$ prevents $e_1$}: If $c_1$ occurs, $e_1$ does not occur ($c_1 \iff \neg e_1$). REQ 3: \enquote{Data redundancy is required to prevent a single failure from causing the loss of collected data.} There will be no data loss due to data redundancy.
\end{compactitem}
\vspace{-.1cm}  
\paragraph{Temporal Ordering of Causes and Effects}
Causes and effects can occur in three different temporal relations~\cite{mostafazadeh16}. In the first temporal relation, the cause occurs before the effect (\emph{before} relation). REQ 1 requires the user to enter a wrong password before the warning window will be displayed. In this example, the cause and effect represent two punctual events. In the second temporal relation, the occurrence of the cause and effect overlaps: \enquote{The fire is burning down the house.} In this case, the occurrence of the emerging fire overlaps with the occurrence of the increasingly brittle house (\emph{overlaps} relation). In the third temporal relation (\emph{during} relation), cause and effect occur simultaneously. REQ 2 describes such a relation, as the effect that you are allowed to do sports on the campus is only valid as long as you have the student status. The start and end time of the cause is therefore also the start and end of the effect. Here, both events are durative.
\vspace{-.1cm}  
\paragraph{Forms of Causality} 
Causality can be expressed in different forms~\cite{blanco08}: \textit{marked} and \textit{unmarked} causality, \textit{explicit} and \textit{implicit} causality, and \textit{ambiguous} and \textit{non-ambiguous} cue phrases.

\begin{compactitem}
    \item \textbf{Marked and unmarked}: A causal relation is \textit{marked} if a certain cue phrase indicates causality. The requirement \enquote{If the user presses the button, a window appears} is \textit{marked} by the cue phrase \enquote{if}, while \enquote{The user has no admin rights. He cannot open the folder.} is \textit{unmarked}.
    \item \textbf{Explicit and implicit}: An \textit{explicit} causal relation provides information about both the cause and effect. The requirement \enquote{In case of an error, the systems prints an error message to the console} is \textit{explicit} as it contains the cause (error) and effect (error message). \enquote{A parent process kills a child process} is \textit{implicit} because the effect that the child process is terminated is not explicitly stated.
    \item \textbf{Ambiguous and non-ambiguous cue phrases}: Given the difference between \textit{marked} and \textit{unmarked} causality, it seems feasible to deduce the presence of causality in a sentence from the occurrence of certain cue phrases. However, there are cue phrases (e.g. since) that may indicate causality, but also occur in other contexts (e.g. to denote time constraints). Such cue phrases are called \textit{ambiguous}, while cue phrases (e.g. because) that mostly indicate causality are called \textit{non-ambiguous}.
\end{compactitem}
\vspace{-.1cm}  
\paragraph{Complexity of Causality}
Our previous explanations refer to the simplest case where the causal relation consists of a single cause and effect. With increasing system complexity, however, the expected system behaviour is described by multiple causes and effects that are connected to each other. They are linked either by conjunctions ($c_1 \land c_2 \land \dots \iff e_1$) or disjunctions ($c_1 \lor c_2 \lor \dots \iff e_1$) or a combination of both which increases the complexity of the causal relation. Furthermore, causal relations can not only be contained in a single sentence, but also span over multiple sentences, which is a significant challenge for causality extraction. Additionally, the complexity increases when several causal relations are linked together, i.e. if the effect of a relation $r_1$ represents a cause in another relation $r_2$. We define such causal relations, where $r_2$ is dependent on $r_1$, as \textit{event chains} (e.g. $r_1: c_1 \iff e_1$ and $r_2: e_1 \iff e_2$).
\vspace{-.3cm} 
\section{Case Study: Causality in Requirement Documents}
\vspace{-.1cm} 
The case study was performed according to the guidelines of Runeson and Höst~\cite{Runeson09}. Based on the classification of Robson~\cite{Robson2002}, our case study is exploratory as we seek for new insights into causality in requirement documents. In this section, we describe our \textit{research questions}, \textit{study objects}, \textit{study design}, \textit{study results}, and \textit{threats to validity}. We also give an overview of the \textit{implications of the study} on the causality detection and extraction from requirements.
\vspace{-.3cm} 
\subsection{Research Questions}
\vspace{-.1cm} 
We are interested in the form and complexity of causality in requirement documents. Based on the terminology introduced in Section~\ref{terminology}, we investigate the following research questions (RQ):

\begin{compactitem}
  \item \textbf{RQ 1}: To which degree does causality occur in requirement documents?
  \item \textbf{RQ 2}: How often do the relations \textit{cause}, \textit{enable} and \textit{prevent} occur?
  \item \textbf{RQ 3}: How often do the temporal relations \textit{before}, \textit{overlap} and \textit{during} occur?
  \item \textbf{RQ 4}: In which form does causality occur in requirement documents? \\ RQ 4a: How often does \textit{marked} and \textit{unmarked} causality occur? \\ RQ 4b: How often does \textit{explicit} and \textit{implicit} causality occur? \\ RQ 4c: Which causal cue phrases are used? Are they mainly \textit{ambiguous} or \textit{non-ambiguous}?
  \item \textbf{RQ 5}: At which complexity does causality occur in requirement documents? \\
  RQ 5a: How often do multiple causes occur? \\ 
  RQ 5b: How often do multiple effects occur? \\ 
  RQ 5c: How often does two sentence causality occur? \\
  RQ 5d: How often do \textit{event chains} occur?
\end{compactitem}
\vspace{-.3cm} 
\subsection{Study Objects}
\begin{figure*}
   \vspace{-.8cm}
    \includegraphics[width=\textwidth]{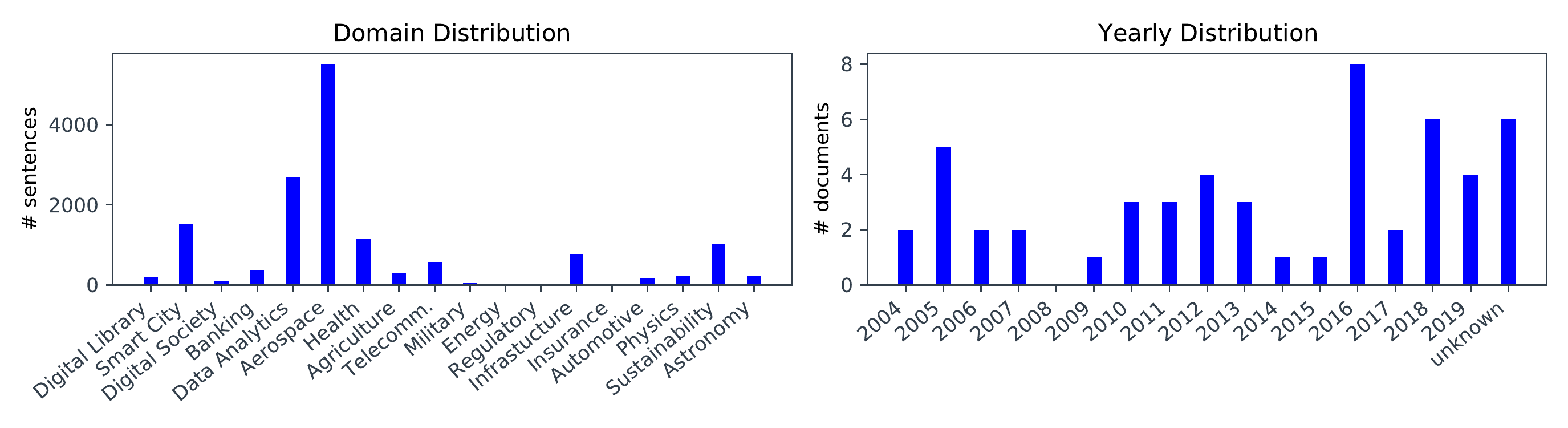}
    \vspace{-.8cm}
  \caption{Descriptive statistics of our data set. The left graph shows the number of sentences per domain. The right graph depicts the year of creation per document.}
  \label{fig:dataset}
  \vspace{-1.0cm}
\end{figure*}

\vspace{-.1cm}
We considered three criteria when selecting a suitable data set for our case study: 1) the data set shall contain requirements documents that are/were used in practice, 2) the data set shall not be domain-specific, rather it shall contain documents from different domains, and 3) the documents shall originate from different years. Consequently, our analysis is not restricted to a single year or domain, but rather allows for a comprehensive view on causality in requirements. Based on these criteria, we selected the data set provided by Fischbach et al.~\cite{fischbach2020}. To the best of our knowledge, this data set is currently the most extensive collection of requirements available in the RE community. It contains 463 documents, from which the authors extracted and pre-processed 212k sentences. For our analysis, we have randomly selected 53 documents from the data set. Our final data set consists of 14,983 sentences from 18 different domains (see Fig.~\ref{fig:dataset}). 

\subsection{Study Design}\label{studyDesign}
\vspace{-.1cm}
\paragraph{Model the phenomenon}
In order to answer our RQ, we need to annotate the sentences in our data set with respect to certain categories (e.g. \textit{explicit} or \textit{implicit} causality). According to Pustejovsky and Stubbs~\cite{Pustejovsky12}, the first step in each annotation process is to \enquote{model the phenomenon} that needs to be annotated. Specifically, it should be defined as a model \emph{M} that consists of a vocabulary \emph{T}, the relations \emph{R} between the terms as well as the interpretations \emph{I} of terms. RQ 1 can be understood as a binary annotation problem, which can be modeled as:

\begin{compactitem}
\item \textbf{T}: \{sentence, causal, not causal\}
\item \textbf{R}: \{sentence ::= causal $\mid$ not causal\}
\item \textbf{I}: \{causal = “A sentence is causal if it contains a relation between at least two events, where e1 causes the occurrence of e2”, $\neg\text{causal}$ = “A sentence is not causal if it describes a state that is independent on any events"\}
\end{compactitem}

Modeling an annotation problem has two advantages: It contributes to a clear definition of the research problem and can be used as a guide for the annotators to explain the meaning of the labels. We have modeled each RQ and discussed it with the annotators. In addition to interpretation \emph{I}, we have also provided an example for each label to avoid misunderstandings. After modeling all RQs, the following nine categories emerged, according to which we annotated our data set: \autour{Causality}, \autour{Explicit}, \autour{Marked}, \autour{Single Sentence}, \autour{Single Cause}, \autour{Single Effect}, \autour{Event Chain}, \autour{Relationship} and \autour{Temporality}.
\vspace{-.1cm} 
\paragraph{Annotation Environment}
We developed our own annotation platform tailored to our research questions.\footnote[2]{The platform can be accessed at \url{clabel.diptsrv003.bth.se}.} Contrary to other annotation platforms~\cite{neves19} which only show single sentences to the annotators, we also show the predecessor and successor of each sentence. This is required to determine whether the causality extends over one sentence or across multiple ones (see RQ 5c). For the binary annotation problems (see RQ 1, RQ 4a, RQ 4b, RQ 5a - d), we provide two labels for each category. Cue phrases present in the sentence can either be selected by the annotator from a list of already labeled cue phrases or new cue phrases can be added using a text input field (see RQ 4c). Since RQ 2 and RQ 3 are ternary annotation problems, the platform provides three labels for these categories. 
\vspace{-.3cm} 
\paragraph{Annotation Guideline} 
Prior to the labeling process, we conducted a workshop with all annotators to ensure a common understanding of causality. The results of the workshop were recorded in the form of an annotation guideline. All annotators were instructed to observe the following annotation rules: First, you should not just check for cue phrases and label the sentence directly as causal, but rather read the sentence completely before making a labeling decision. Otherwise, too many False Positives will be introduced.
Second, you should check if the cause is really necessary for the effect to occur. Only if the cause is mandatory for the effect, it is a causal relation. 

\begin{table}
\vspace{-.5cm}
\centering
\setlength{\extrarowheight}{0pt}
\addtolength{\extrarowheight}{\aboverulesep}
\addtolength{\extrarowheight}{\belowrulesep}
\setlength{\aboverulesep}{0pt}
\setlength{\belowrulesep}{0pt}
\caption{Inter-annotator agreement statistics per category. The two categories Relationship and Temporality were jointly labeled by the first and second author and therefore do not require a reliability assessment.}
\label{Tab:measures}
\vspace{-.2cm}
\resizebox{\textwidth}{!}{\begin{tabular}{llcc|cc|cc|cc|cc|cc|cc|c} 
\toprule
                       &                  & \multicolumn{2}{c|}{\thead{Causal } }                                             & \multicolumn{2}{c|}{\thead{Explicit } }                                         & \multicolumn{2}{c|}{\thead{Marked } }                                           & \multicolumn{2}{c|}{\thead{Single \\ Sentence} }                                   & \multicolumn{2}{c|}{\thead{Single \\ Cause } }                                     & \multicolumn{2}{c|}{\thead{Single \\ Effect } }                                    & \multicolumn{2}{c|}{\thead{Event \\ Chain } }                                      & \textbf{avg.}                         \\
\vcell{}               & \vcell{}         & \vcell{\textbf{0} }                      & \vcell{\textbf{1} }                     & \vcell{\textbf{0} }                    & \vcell{\textbf{1} }                     & \vcell{\textbf{0} }                    & \vcell{\textbf{1} }                     & \vcell{\textbf{0} }                    & \vcell{\textbf{1} }                     & \vcell{\textbf{0} }                    & \vcell{\textbf{1} }                     & \vcell{\textbf{0} }                    & \vcell{\textbf{1} }                     & \vcell{\textbf{0} }                     & \vcell{\textbf{1} }                    & \multicolumn{1}{l}{\vcell{}}          \\[-\rowheight]
\printcellbottom       & \printcellmiddle & \printcellmiddle                         & \printcellmiddle                        & \printcellmiddle                       & \printcellmiddle                        & \printcellmiddle                       & \printcellmiddle                        & \printcellmiddle                       & \printcellmiddle                        & \printcellmiddle                       & \printcellmiddle                        & \printcellmiddle                       & \printcellmiddle                        & \printcellmiddle                        & \printcellmiddle                       & \multicolumn{1}{l}{\printcellmiddle}  \\
\textbf{Confusion}     & \textbf{0}       & {\cellcolor[rgb]{0.753,0.753,0.753}}2034 & {\cellcolor[rgb]{0.753,0.753,0.753}}193 & {\cellcolor[rgb]{0.753,0.753,0.753}}24 & {\cellcolor[rgb]{0.753,0.753,0.753}}25  & {\cellcolor[rgb]{0.753,0.753,0.753}}1  & {\cellcolor[rgb]{0.753,0.753,0.753}}22  & {\cellcolor[rgb]{0.753,0.753,0.753}}12 & {\cellcolor[rgb]{0.753,0.753,0.753}}8   & {\cellcolor[rgb]{0.753,0.753,0.753}}41 & {\cellcolor[rgb]{0.753,0.753,0.753}}77  & {\cellcolor[rgb]{0.753,0.753,0.753}}63 & {\cellcolor[rgb]{0.753,0.753,0.753}}72  & {\cellcolor[rgb]{0.753,0.753,0.753}}450 & {\cellcolor[rgb]{0.753,0.753,0.753}}27 &                                       \\
\textbf{Matrix}        & \textbf{1}       & {\cellcolor[rgb]{0.753,0.753,0.753}}274  & {\cellcolor[rgb]{0.753,0.753,0.753}}499 & {\cellcolor[rgb]{0.753,0.753,0.753}}39 & {\cellcolor[rgb]{0.753,0.753,0.753}}411 & {\cellcolor[rgb]{0.753,0.753,0.753}}12 & {\cellcolor[rgb]{0.753,0.753,0.753}}464 & {\cellcolor[rgb]{0.753,0.753,0.753}}17 & {\cellcolor[rgb]{0.753,0.753,0.753}}462 & {\cellcolor[rgb]{0.753,0.753,0.753}}43 & {\cellcolor[rgb]{0.753,0.753,0.753}}338 & {\cellcolor[rgb]{0.753,0.753,0.753}}46 & {\cellcolor[rgb]{0.753,0.753,0.753}}318 & {\cellcolor[rgb]{0.753,0.753,0.753}}13  & {\cellcolor[rgb]{0.753,0.753,0.753}}9  & \multicolumn{1}{l}{}                  \\
\textbf{Agreement}     &                  & \multicolumn{2}{c|}{84.4~\%}                                                        & \multicolumn{2}{c|}{87.2~\%}                                                      & \multicolumn{2}{c|}{93.1~\%}                                                      & \multicolumn{2}{c|}{95.0~\%}                                                      & \multicolumn{2}{c|}{76.0~\%}                                                      & \multicolumn{2}{c|}{76.4~\%}                                                      & \multicolumn{2}{c|}{92.0~\%}                                                      & 86.3~\%                                \\
\textbf{Cohen's Kappa} &                  & \multicolumn{2}{c|}{0.579}                                                         & \multicolumn{2}{c|}{0.358}                                                       & \multicolumn{2}{c|}{0.023}                                                       & \multicolumn{2}{c|}{0.464}                                                       & \multicolumn{2}{c|}{0.261}                                                       & \multicolumn{2}{c|}{0.362}                                                       & \multicolumn{2}{c|}{0.27}                                                        & 0.331                                 \\
\textbf{Gwet's AC1}    &                  & \multicolumn{2}{c|}{0.753}                                                         & \multicolumn{2}{c|}{0.84}                                                        & \multicolumn{2}{c|}{0.926}                                                       & \multicolumn{2}{c|}{0.945}                                                       & \multicolumn{2}{c|}{0.645}                                                       & \multicolumn{2}{c|}{0.625}                                                       & \multicolumn{2}{c|}{0.91}                                                        & 0.806                                 \\
\bottomrule
\end{tabular}}
\vspace{-1.0cm}
\end{table}

\paragraph{Annotation Validity}
To verify the reliability of our annotations, we calculated the inter-annotator agreement. We assigned 3,000 sentences to each annotator, of which 2,500 are unique and 500 overlapping. Based on the overlapping sentences, we calculated the Cohen's Kappa~\cite{cohen60} measure to evaluate how well the annotators can make the same annotation decision for a given category. We chose Cohen's Kappa since it is widely used for assessing inter-rater reliability~\cite{Viera05}. However, a number of statistical problems are known to exist with this measure~\cite{McHugh12}. In case of a high imbalance of ratings, Cohen's Kappa is low and indicates poor inter-rater reliability even if there is a high agreement between the raters (Kappa paradox~\cite{FEINSTEIN1990}). Thus, Cohen's Kappa is not meaningful in such scenarios. Consequently, studies~\cite{Wongpakaran13} suggest that Cohen's Kappa should always be reported together with the percentage of agreement and other paradox resistant measures (e.g. Gwet's AC1 measure~\cite{gwet}) in order to make a valid statement about the inter-rater reliability. We involved six annotators in the creation of the corpus and assessed the inter-rater reliability on the basis of 3,000 overlapping sentences, which represents about 20~\% of the total data set. We calculated all measures (see Tab.~\ref{Tab:measures}) using the cloud-based version of AgreeStat~\cite{AgreeStat}. Cohen's Kappa and Gwet's AC1 can both be interpreted using the taxonomy developed by Landis and Koch~\cite{landis77}: values $\leq$ 0 as indicating no agreement and 0.01–0.20 as none to slight, 0.21–0.40 as fair, 0.41–0.60 as moderate, 0.61–0.80 as substantial, and 0.81–1.00 as almost perfect agreement. Tab.~\ref{Tab:measures} demonstrates that the inter-rater agreement of our annotation process is reliable. Across all categories, an average percentage of agreement of 86~\% was achieved. Except for the categories \autour{Single Cause} and \autour{Single Effect}, all categories show a percentage of agreement of at least 84~\%. We hypothesize that the slightly lower value of 76~\% for these two categories is caused by the fact that in some cases the annotators interpret the causes and effects with different granularity (e.g., annotators might break some causes and effects down into several sub causes and effects, while some do not). Hence, the annotations  differ slightly. The Kappa paradox is particularly evident for the categories \autour{Marked} and \autour{Event Chain}. Despite a high agreement of over 90~\%, Cohen's Kappa yields a very low value, which \enquote{paradoxically} suggests almost no or only fair agreement. A more meaningful assessment is provided by Gwet's AC1 as it did not fail in case of prevalence and remains close to the percentage of agreement. Across all categories, the mean value is above 0.8, which indicates a nearly perfect agreement. Therefore, we assess our labeled data set as reliable and suitable for further analysis and the implementation of our causality detection approach.

\begin{figure*}
\vspace{-.5cm}
\centering
    \includegraphics[width=\textwidth]{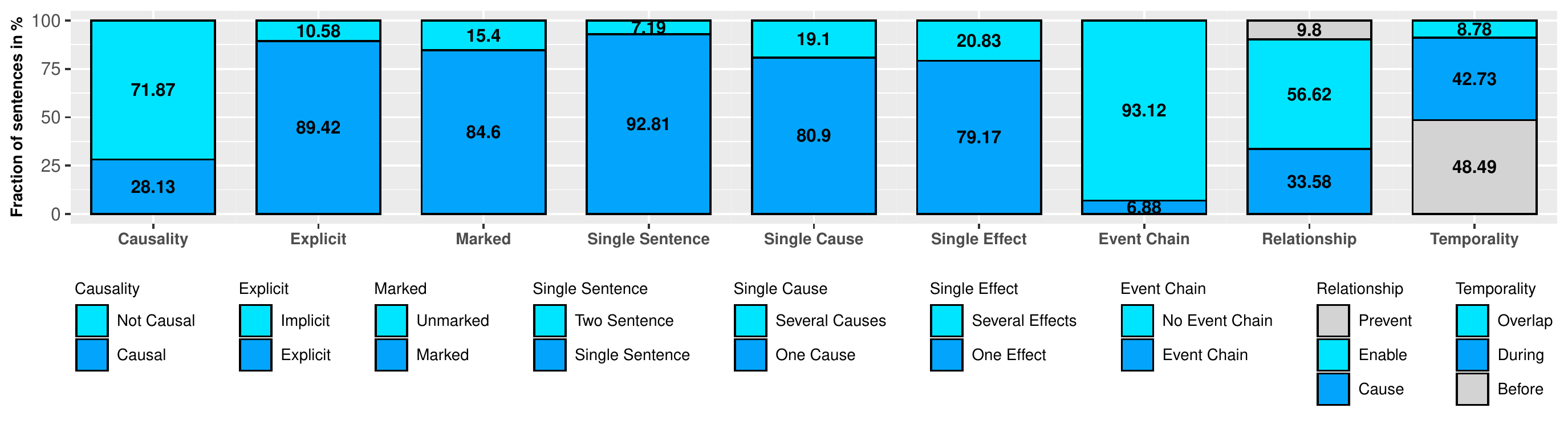}
        \vspace{-.9cm}
  \caption{Annotation results per category. The y axis of the bar plot for the category \enquote{Causality} refers to the total number of analyzed sentences. The other bar plots are only related to the causal sentences.}
  \label{fig:studyResults}
      \vspace{-.9cm}
\end{figure*}

\subsection{Study Results}
\vspace{-.1cm}
Fig.~\ref{fig:studyResults} presents the analysis results for each labeled category. When interpreting the values, it is important to note that we analyze entire requirement documents in our study. Consequently, our data set contains records with different contents, which do not necessarily represent all functional requirements. For example, requirement documents also contain non-functional requirements, phrases for content structuring, purpose statements, etc. Hence, the results of our analysis do not only refer to functional requirements but in general to the content of requirement documents.

\textbf{Answer to RQ1:} Fig.~\ref{fig:studyResults} highlights that causality occurs in requirement documents. About 28~\% of the analyzed sentences are causal. It can therefore be concluded that causality is a major linguistic element of requirement documents since almost one third of all sentences are causal.

\textbf{Answer to RQ2:} 
The majority (56~\%) of causal sentences contained in requirement documents express an \textit{enable} relationship between certain events. Only about 10~\% of the causal sentences indicate a \textit{prevent} relationship. \textit{Cause} relationships are found in about 34~\% of the annotated data.

\textbf{Answer to RQ3:}
Interestingly, we found that causes and effects occur almost equally often in a \textit{before} and \textit{during} relation. With about 48~\%, the \textit{before} relation is the most frequent temporal relation in our data set, but only with a difference of about 6~\% compared to the \textit{during} relation. The \textit{overlap} relation occurred only in a minority (8.78~\% of the sentences).

\textbf{Answer to RQ4a:} Fig.~\ref{fig:studyResults} shows that the majority of causal sentences contain one or more cue phrases to indicate the causal relationship between certain events. \textit{Unmarked} causality occurs only in about 15~\% of the analyzed sentences.

\textbf{Answer to RQ4b:} Most causal sentences are \textit{explicit}, i.e. they contain information about both the cause and the effect. Only about 10~\% of causal sentences are \textit{implicit}.

\textbf{Answer to RQ4c:} 
Tab.~\ref{Tab:cuePhrases} provides an overview of the causal cue phrases used in the requirement documents. The left side of the table shows the different cue phrases ordered by word group. On the right side, all verbs used to express causal relations are listed. We order the verbs according to whether they express a \textit{cause}, \textit{enable} or \textit{prevent} relationship. To measure the ambiguity of the individual cue phrases, we introduce the ambiguity factor (AF). We define AF for a cue phrase x as the conditional probability that a sentence is causal given that the cue phrase x occurs in the sentence: $\Pr( \text{Causal} \mid \text{X is present in sentence})$. Hence, a high AF value indicates a \textit{non-ambiguous} cue phrase, while low values indicate strongly \textit{ambiguous} cue phrases. Tab.~\ref{Tab:cuePhrases} demonstrates that a number of different cue phrases are used to express causality in requirement documents. Not surprisingly, cue phrases like \enquote{if}, \enquote{because} and \enquote{therefore} show AF values of more than 90~\%. However, there is a variety of cue phrases that indicate causality in some sentences but also occur in other non-causal contexts. This is especially evident in the case of pronouns. Relative sentences can indicate causality, but not in every case, which is reflected by the low AF value. A similar pattern emerges with regard to the used verbs. Only a few verbs (e.g., \enquote{leads to, degrade and enhance}) show a high AF value. Consequently, the majority of used verbs do not necessarily indicate a causal relation if they are present in a sentence. 

\textbf{Answer to RQ 5a:} Fig.~\ref{fig:studyResults} illustrates that a causal relation in requirement documents often includes only a single cause. Multiple causes occur in only 19.1~\% of analyzed causal sentences. The exact number of causes was not documented during the annotation process. However, the participating annotators reported consistently that in the case of complex causal relations, two to three causes were usually included. More than three causes were rare. 

\textbf{Answer to RQ5b:} Interestingly, the distribution of effects is similar to that of causes. Likewise, single effects occur significantly more often than multiple effects. According to the annotators, the number of effects in case of complex relations is limited to two effects. Three or more effects occur rarely.

\textbf{Answer to RQ5c:} Most causal relations can be found in single sentences. Relations where cause and effect are distributed over several sentences occur only in about 7~\% of the analyzed data. The annotators reported that most often the cue phrase \enquote{therefore} was used to express two-sentence causality.

\textbf{Answer to RQ5d:} Fig.~\ref{fig:studyResults} shows that \textit{event chains} are rarely used in requirement documents. Most causal sentences contain isolated causal relations and only a few \textit{event chains}.

\begin{table}[ht]
\centering
\caption{Overview of cue phrases used to indicate causality in requirement documents. \textbf{Bold} AF values highlight non-ambiguous phrases that mostly indicated causality ($\Pr( \text{Causal} \mid \text{X is present in sentence}) \geq 0.8$).}
\vspace{-.1cm}
\resizebox{\textwidth}{!}{\begin{tabular}{lllll||lllll} 
\toprule
\textbf{Type} & \textbf{Phrase}   & \textbf{Causal} & \textbf{Not Causal} & \begin{tabular}[c]{@{}l@{}}\textbf{Ambiguity }\\\textbf{Factor (AF)}\end{tabular} & \textbf{Type} & \textbf{Phrase}   & \textbf{Causal} & \textbf{Not Causal} & \textbf{AF}    \\ 
\midrule
conjunctions  & if                & 387             & 41                  & \textbf{0.90}                                                                     & Cause         & force(s/ed)       & 21              & 18                  & 0.54           \\
              & as                & 607             & 1313                & 0.32                                                                              &               & cause(s/ed)       & 32              & 10                  & 0.76           \\
              & because           & 78              & 7                   & \textbf{0.92}                                                                     &               & lead(s) to        & 5               & 0                   & \textbf{1.00}  \\
              & but               & 100             & 204                 & 0.33                                                                              &               & reduce(s/ed)      & 48              & 28                  & 0.63           \\
              & in order to       & 141             & 33                  & \textbf{0.81}                                                                     &               & minimize(s/ed)    & 28              & 11                  & 0.72           \\
              & so (that)         & 88              & 86                  & 0.51                                                                              &               & affect(s/ed)      & 13              & 19                  & 0.41           \\
              & unless            & 23              & 4                   & \textbf{0.85}                                                                     &               & maximize(s/ed)    & 11              & 5                   & 0.69           \\
              & while             & 71              & 90                  & 0.44                                                                              &               & eliminate(s/ed)   & 8               & 11                  & 0.42           \\
              & once              & 48              & 15                  & 0.76                                                                              &               & result(s/ed) in   & 50              & 43                  & 0.54           \\
              & except            & 9               & 5                   & 0.64                                                                              &               & increase(s/ed)    & 49              & 34                  & 0.59           \\
              & as long as        & 12              & 1                   & \textbf{0.92}                                                                     &               & decrease(s/ed)    & 5               & 8                   & 0.38           \\ 
\cline{1-5}
adverbs       & therefore         & 61              & 6                   & \textbf{0.91}                                                                     &               & impact(s)         & 37              & 68                  & 0.35           \\
              & when              & 331             & 64                  & \textbf{0.84}                                                                     &               & degrade(s/ed)     & 11              & 2                   & \textbf{0.85}  \\
              & whenever          & 10              & 0                   & \textbf{1.00}                                                                     &               & introduce(s/ed)   & 11              & 12                  & 0.48           \\
              & hence             & 21              & 9                   & 0.70                                                                              &               & enforce(s/ed)     & 2               & 1                   & 0.67           \\
              & where             & 213             & 150                 & 0.59                                                                              &               & trigger(s/ed)     & 11              & 7                   & 0.61           \\ 
\cline{6-10}
              & since             & 65              & 32                  & 0.67                                                                              & Enable        & depend(s) on      & 28              & 21                  & 0.57           \\
              & consequently      & 2               & 6                   & 0.25                                                                              &               & require(s/ed)     & 316             & 262                 & 0.55           \\
              & wherever          & 5               & 2                   & 0.71                                                                              &               & allow(s/ed)       & 187             & 130                 & 0.59           \\
              & rather            & 16              & 30                  & 0.35                                                                              &               & need(s/ed)        & 98              & 162                 & 0.38           \\
              & to this/that end  & 12              & 0                   & \textbf{1.00}                                                                     &               & necessitate(s/ed) & 7               & 2                   & 0.78           \\
              & thus              & 66              & 17                  & \textbf{0.80}                                                                     &               & facilitate(s/ed)  & 29              & 28                  & 0.51           \\
              & for this reason   & 7               & 3                   & 0.70                                                                              &               & enhance(s/ed)     & 16              & 4                   & \textbf{0.80}  \\
              & due to            & 91              & 26                  & 0.78                                                                              &               & ensure(s/ed)      & 145             & 66                  & 0.69           \\
              & thereby           & 4               & 2                   & 0.67                                                                              &               & achieve(s/ed)     & 30              & 24                  & 0.56           \\
              & as a result       & 11              & 4                   & 0.73                                                                              &               & support(s/ed)     & 128             & 301                 & 0.30           \\
              & for this purpose  & 1               & 2                   & 0.33                                                                              &               & enable(s/ed)      & 75              & 36                  & 0.68           \\ 
\cline{1-5}
pronouns      & which             & 277             & 608                 & 0.31                                                                              &               & permit(s/ed)      & 10              & 13                  & 0.43           \\
              & who               & 28              & 52                  & 0.35                                                                              &               & rely on           & 3               & 5                   & 0.38           \\ 
\cline{6-10}
              & that              & 732             & 1178                & 0.38                                                                              & Prevent       & hinder(s/ed)      & 1               & 1                   & 0.50           \\
              & whose             & 16              & 11                  & 0.59                                                                              &               & prevent(s/ed)     & 38              & 17                  & 0.69           \\ 
\cline{1-5}
adjectives    & only              & 127             & 126                 & 0.50                                                                              &               & avoid(s/ed)       & 14              & 23                  & 0.38           \\
              & prior to          & 26              & 20                  & 0.57                                                                              &               &                   &                 &                     &                \\
              & imperative        & 1               & 3                   & 0.25                                                                              &               &                   &                 &                     &                \\
              & necessary (to)    & 36              & 19                  & 0.65                                                                              &               &                   &                 &                     &                \\ 
\cline{1-5}
preposition   & for               & 1209            & 2753                & 0.31                                                                              &               &                   &                 &                     &                \\
              & during            & 327             & 137                 & 0.70                                                                              &               &                   &                 &                     &                \\
              & after             & 133             & 57                  & 0.70                                                                              &               &                   &                 &                     &                \\
              & by                & 506             & 1171                & 0.30                                                                              &               &                   &                 &                     &                \\
              & with              & 680             & 1554                & 0.30                                                                              &               &                   &                 &                     &                \\
              & in the course of  & 2               & 1                   & 0.67                                                                              &               &                   &                 &                     &                \\
              & through           & 114             & 204                 & 0.36                                                                              &               &                   &                 &                     &                \\
              & as part of        & 19              & 51                  & 0.27                                                                              &               &                   &                 &                     &                \\
              & in this case      & 18              & 3                   & \textbf{0.86}                                                                     &               &                   &                 &                     &                \\
              & before            & 54              & 27                  & 0.67                                                                              &               &                   &                 &                     &                \\
              & until             & 33              & 11                  & 0.75                                                                              &               &                   &                 &                     &                \\
              & upon              & 25              & 48                  & 0.34                                                                              &               &                   &                 &                     &                \\
              & in case of        & 30              & 7                   & \textbf{0.81}                                                                     &               &                   &                 &                     &                \\
              & in both cases     & 1               & 0                   & \textbf{1.00}                                                                     &               &                   &                 &                     &                \\
              & in the event of   & 15              & 2                   & \textbf{0.88}                                                                     &               &                   &                 &                     &                \\
              & in response to    & 6               & 7                   & 0.46                                                                              &               &                   &                 &                     &                \\
              & in the absence of & 8               & 1                   & \textbf{0.89}                                                                     &               &                   &                 &                     &                \\
\bottomrule
\end{tabular}}
\label{Tab:cuePhrases}
\vspace{-.5cm}
\end{table}

\subsection{Implications for Causality Detection and Extraction}
Based on the results of our case study, we draw the following conclusions: Causality matters in requirements documents, which underlines the necessity of an approach for the automatic detection and extraction of causal requirements. The complexity of causal relations ranges from low to medium, since they usually consist of a single cause and effect relationship. However, for the approaches to be applicable in practice, they need to comprehend also more complex relations containing between two to three causes and two effects. Hence, the approaches must be capable of understanding conjunctions, disjunctions and negations in the sentences to fully capture the relationships between causes and effects. Two-sentence causality and \textit{event chains} occur only rarely. Thus, both aspects can initially be neglected in the development of the approaches, while still more than 92~\% of the analyzed sentences can be covered. Since most causal relations in requirements documents are \textit{explicit}, the detection and extraction of causality is simplified. The information about both causes and effects is embedded directly in the sentences, so that the approaches require little or no \textit{implicit} knowledge. The analysis of the AF values reveals that most of the used cue phrases are ambiguous. Consequently, our methods require a deep understanding of language as causality can not only be deduced from the presence of certain cue phrases but rather from a combination of the syntax and semantics of the sentence.
\vspace{-.3cm}
\subsection{Threats to Validity}
\vspace{-.1cm}
\textit{Internal Validity}: A major threat to internal validity are the annotations themselves as an annotation task is to a certain degree subjective. To minimize the bias of the annotators, we performed two mitigation actions: First, we conducted a workshop prior to the annotation process to ensure a common understanding of causality. Second, we assessed the inter-rater agreement by using multiple metrics (Gwet's AC1 etc.).
\textit{External Validity}: To achieve reasonable generalizability, we selected requirements documents from different domains and years. As Fig.~\ref{fig:dataset} shows, our data set covers a variety of domains, but the distribution of the sentences is imbalanced. The domains aerospace, data analytics, and smart city account for a large part of the data set (9,724 sentences), while the other 15 domains are underrepresented. Hence, our results do not allow a general conclusion about causality in requirements documents. Future studies should expand to more documents from these underrepresented as well as further domains to achieve a more global insight into causality in requirements documents.
\vspace{-.4cm}
\section{Approach: Detecting Causal Requirements}
\vspace{-.2cm}
This section presents the implementation of our causal classifier. Initially, we describe our applied methods followed by a report of the results of our experiments, in which we compare the performance of the individual methods.
\vspace{-.4cm}
\subsection{Methods}
\vspace{-.2cm}
\paragraph{\textbf{Rule Based Approach}}
The baseline approach for causality detection involves the use of simple regex expressions. We iterate through all sentences in the test set and check if one of the phrases listed in Tab.~\ref{Tab:cuePhrases} is included. For the positive case, the sentence is classified as causal and vice versa.
\vspace{-.2cm}
\paragraph{\textbf{Machine Learning Based Approach}}
As a second approach, we investigate the use of \textit{supervised} ML models that learn to predict causality based on the labeled data set. Specifically, we employ established binary classification algorithms: Naive Bayes (NB), Support Vector Machines (SVM), Random Forest (RF), Decision Tree (DT), Logistic Regression (LR), Ada Boost (AB) and K-Nearest Neighbor (KNN). To determine the best hyperparameters for each binary classifier, we apply Grid Search, which fits the model on every possible combination of hyperparameters and selects the most performant. We use two different methods as word embeddings: Bag of Words (BoW) and Term Frequency-Inverse Document Frequency (TF-IDF). In Tab.~\ref{Tab:experimentalResults} we report the classification results of each algorithm as well as the best combination of hyperparameters.
\vspace{-.2cm}
\paragraph{\textbf{Deep Learning Based Approach}}
With the rise of Deep Learning (DL), more and more researchers are using DL models for Natural Language Processing (NLP) tasks. In this context, the Bidirectional Encoder Representations from Transformers (BERT) model~\cite{devlin19} is prominent and has already been used for question answering and named entity recognition. BERT is pre-trained on large corpora and can therefore easily be fine tuned for any downstream task without the need for much training data (Transfer Learning). In our paper, we make use of the fine tuning mechanism of BERT and investigate to which extent it can be used for causality detection of requirement sentences. First, we tokenize each sentence. BERT requires input sequences with a fixed length (maximum 512 tokens). Therefore, for sentences that are shorter than this fixed length, padding tokens (PAD) are inserted to adjust all sentences to the same length. Other tokens, such as the classification (CLS) token, are also inserted in order to provide further information of the sentence to the model. CLS is the first token in the sequence and represents the whole sentence (i.e. it is the pooled output of all tokens of a sentence). For our classification task, we mainly use this token because it stores the information of the whole sentence. We feed the pooled information into a single-layer feedforward neural network that uses a softmax layer, which calculates the probability that a sentence is causal or not. We tune BERT in three different ways and investigate their performance:

\begin{compactitem}
  \item \textbf{BERT\textsubscript{Base}} In the base variant, the sentences are tokenized as described above and put into the classifier. To choose a suitable fixed length for our input sequences, we analyzed the lengths of the sentences in our data set. Even with a fixed length of 128 tokens we cover more than 97~\% of the sentences. Sentences containing more tokens are shortened accordingly. Since this is only a small amount, only little information is lost. Thus, we chose a fixed length of 128 tokens instead of the maximum possible 512 tokens to keep BERT's computational requirements to a minimum. 
  \item \textbf{BERT\textsubscript{POS}} Studies have shown that the performance of NLP models can be improved by providing explicit prior knowledge of syntactic information to the model~\cite{sundararaman2019}. Therefore, we enrich the input sequence with syntactic information and feed it into BERT. More specifically, we add the corresponding Part-of-speech (POS) tag to each token by using the spaCy NLP library~\cite{spacy2}. One way to encode the input sequence with the corresponding POS tags is to concatenate each token embedding with a hot encoded vector representing the POS tag. Since the BERT token embeddings are high dimensional, the impact of a single added feature (i.e. the POS tag) would be low. Contrary, we hypothesize that the syntactic information has a higher impact if we annotate the input sentences directly with the POS tags and then put the annotated sentences into BERT. This way of creating linguistically enriched input sequences has already proven to be promising during the development of the NLPL word embeddings~\cite{fares17}. Fig.~\ref{fig:inputSequence} shows how we incorporated the POS tags into the input sequence. By extending the input sequence, the fixed length for the BERT model has to be adapted accordingly. After a further analysis, a length of 384 tokens proved to be reasonable.
  \item \textbf{BERT\textsubscript{DEP}} Similar to the previous fine-tuning approach, we follow the idea of enriching the input sequence by linguistic features. Instead of using the POS tags, we use the dependency (DEP) tags (see Fig.~\ref{fig:inputSequence}) of each token. Thus, we provide knowledge about the grammatical structure of the sentence to the classifier. We hypothesize that this knowledge has a positive effect on the model performance, as a causal relation is a specific grammatical structure (e.g. it often contains an adverbial clause) and the classifier can learn causal specific patterns in the grammatical structure of the training instances. The fixed token length was also increased to 384 tokens.
\end{compactitem}

\begin{figure}
\vspace{-.5cm}
\begin{mdframed} 
   {\scriptsize \textsc{Bert}\textsubscript{Base}: If the process fails, an error message is shown.}
   
   {\scriptsize \textsc{Bert}\textsubscript{POS}: If\_\blue{SCONJ} the\_\blue{DET} process\_\blue{NOUN} fails\_\blue{VERB} ,\_\blue{PUNCT} an\_\blue{DET} error\_\blue{NOUN} message\_\blue{NOUN} is\_\blue{AUX} shown\_\blue{VERB} .\_\blue{PUNCT}}
   
   {\scriptsize \textsc{Bert}\textsubscript{DEP}: If\_\orange{mark} the\_\orange{det} process\_\orange{nsubj} fails\_\orange{advcl} ,\_\orange{punct} an\_\orange{det} error\_\orange{compound} message\_\orange{nsubjpass} is\_\orange{auxpass} shown\_\orange{ROOT} .\_\orange{punct}}
   \end{mdframed}
       \vspace{-.5cm}
    \caption{Input sequences used for our different BERT fine tuning models. POS tags are marked orange and DEP tags are marked blue.}
    \label{fig:inputSequence}
        \vspace{-.9cm}
\end{figure}

\vspace{-.1cm} 
\subsection{Evaluation Procedure}
\begin{figure*}
\vspace{-.5cm}
\centering
\scalebox{0.9}{
    \includegraphics[width=\textwidth]{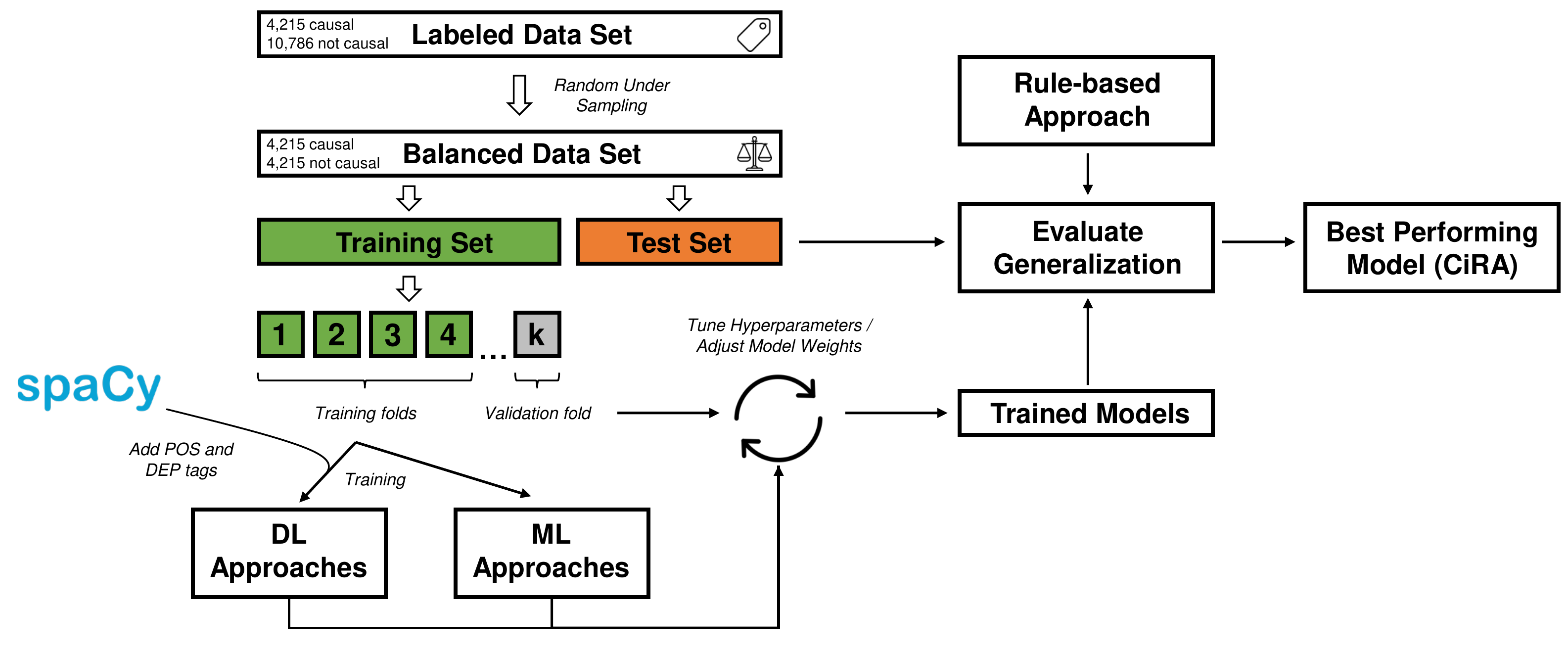}}
    \vspace{-.4cm}
  \caption{Implementation and Evaluation Procedure of our Binary Classifier}
  \label{fig:implementationProcedure}
  \vspace{-.9cm}
\end{figure*}

\vspace{-.1cm}
Our labeled data set is imbalanced as only 28.1~\% are positive samples. To avoid the class imbalance problem, we apply Random Under Sampling (see Fig.~\ref{fig:implementationProcedure}). We randomly select sentences from the majority class and exclude them from the data set until a balanced distribution is achieved. Our final data set consists of 8,430 sentences of which 4,215 are equally causal and non-causal. We follow the idea of Cross Validation and divide the data set in a training, validation and test set. The training set is used for fitting the algorithm while the validation set is used to tune its parameters. The test set is utilized for the evaluation of the algorithm based on real world unseen data. We opt for a 10-fold Cross Validation as a number of studies have shown that a model that has been trained this way demonstrates low bias and variance~\cite{James13}. We use standard metrics, for evaluating our approaches: Accuracy, Precision, Recall and F\textsubscript{1} score~\cite{James13}. When interpreting the metrics, it is important to consider which misclassification (False Negative or False Positive) matters most resp. causes the highest costs. Since causality detection is supposed to be the first step towards automatic causality extraction, we favor Recall over Precision. A high Recall corresponds to a greater degree of automation of causality extraction, because it is easier for users to discard False Positives then to manually detect False Negatives. Consequently, we seek high Recall to minimize the risk of missed causal sentences and acceptable Precision to ensure that users are not overwhelmed by False Positives. 

\subsection{Experimental Results}
\vspace{-.1cm}
Tab.~\ref{Tab:experimentalResults} demonstrates the inability of the baseline approach to distinguish between causal (F\textsubscript{1} score: 66~\%) and non-casual (F\textsubscript{1} score: 64~\%) sentences. This coincides with our observation from the case study that searching for cue phrases is not suitable for causality detection. In comparison, most ML based approaches (except KN and DT) show a better performance. The best performance in this category is achieved by RF with an Accuracy of 78~\% (gain of 13~\% compared to baseline approach). The overall best classification results are achieved by our DL based approaches. All three variants were trained with the hyperparameters recommended by Devlin et al.~\cite{devlin19}. Even the vanilla \textbf{BERT\textsubscript{Base}} model shows a great performance in both classes (F\textsubscript{1} score $\geq$ 80~\% for causal and non-causal). Interestingly, enriching the input sequences with syntactic information did not result in a significant performance boost. \textbf{BERT\textsubscript{POS}} even has a slightly worse Accuracy value of 78~\% (difference of 2~\% compared to \textbf{BERT\textsubscript{Base}}). An improvement of the performance can be observed in the case of \textbf{BERT\textsubscript{DEP}}, which has the best F\textsubscript{1} score for both classes among all the other approaches and also achieves the highest Accuracy value of 82~\%. Compared to the rule based and ML based approaches, \textbf{BERT\textsubscript{DEP}} yields an average gain of 11.06~\% in macro-Recall and 11.43~\% in macro-Precision. Interesting is a comparison with \textbf{BERT\textsubscript{Base}}. \textbf{BERT\textsubscript{DEP}} shows better values across all metrics, but the difference is only marginal. This indicates that \textbf{BERT\textsubscript{Base}} already has a deep language understanding due to its pre-training and therefore can be tuned well for causality detection without much further input. However, over all five runs, the use of the DEP tags shows a small but not negligible performance gain - especially regarding our main decision criterion: the Recall value (85~\% for causal and 79~\% for non-causal). Therefore, we choose \textbf{BERT\textsubscript{DEP}} as our final approach (CiRA).

\begin{table}
\vspace{-.5cm}
\centering
\caption{Recall, Precision, F\textsubscript{1} scores (per class) and Accuracy. We report the averaged scores over five repetitions and highlight in \textbf{bold} the best results for each metric.}
\label{Tab:experimentalResults}
\vspace{-.2cm}
\resizebox{\textwidth}{!}{\begin{tabular}{lllccccccc} 
\toprule
                                   &           &                                                                                                & \multicolumn{3}{c}{\textbf{Causal (Support: 435)}} & \multicolumn{3}{c}{\textbf{Not Causal (Support: 408)}} & \multicolumn{1}{l}{}   \\
                                   &           & Best hyperparameters                                                                           & Recall        & Precision     & F1                 & Recall        & Precision     & F1                     & Accuracy                    \\
\textbf{Rule based}                &           & -                                                                                              & 0.65          & 0.66          & 0.66               & 0.65          & 0.63          & 0.64                   & 0.65                                      \\ 
\hline
\multirow{7}{*}[-5em]{\textbf{ML based}} & NB        &
\begin{tabular}{@{}l@{}}alpha: 1, fit\_prior: True,\\embed: BoW\end{tabular}                                              & 0.71          & 0.7           & 0.71               & 0.68          & 0.69          & 0.69                   & 0.7                                     \\  \addlinespace
                                   & SVM       &
                                   \begin{tabular}{@{}l@{}}C: 50, gamma: 0.001,\\kernel: rbf, embed: BoW\end{tabular}                                                                     & 0.68          & 0.8           & 0.73               & 0.82          & 0.71          & 0.76                   & 0.75                                  \\  \addlinespace
                                   & RF        &
                                   \begin{tabular}{@{}l@{}}criterion: entropy, max\_features: auto,\\ n\_estimators: 500, embed: BoW\end{tabular}
                                   
                                                & 0.72          & \textbf{0.82} & 0.77               & \textbf{0.84} & 0.74          & 0.79                   & 0.78                                  \\  \addlinespace
                                   & DT        &
                                   
                                   \begin{tabular}{@{}l@{}}criterion: gini, max\_features: auto,\\ splitter: random, embed: TF-IDF \end{tabular}

                                                              & 0.65          & 0.68          & 0.66               & 0.67          & 0.65          & 0.66                   & 0.66                                   \\  \addlinespace
                                   & LR        & 
                                   
                                   \begin{tabular}{@{}l@{}} C: 1, solver: liblinear,\\  embed: TF-IDF\end{tabular}

                                                                                           & 0.71          & 0.78          & 0.74               & 0.79          & 0.72          & 0.75                   & 0.75                                  \\  \addlinespace
                                   & AB        &
                                   
                        \begin{tabular}{@{}l@{}}algorithm: SAMME.R, n\_estimators: 200,\\ embed: BoW \end{tabular}           
                                   
                                                                      & 0.67          & 0.78          & 0.72               & 0.8           & 0.7           & 0.75                   & 0.74                                  \\  \addlinespace
                                   & KNN        &
                                   
                            \begin{tabular}{@{}l@{}}algorithm: ball\_tree, n\_neighbors: 20,\\ weights: distance, embed: TF-IDF \end{tabular}        
                                   
                                                           & 0.61          & 0.68          & 0.64               & 0.7           & 0.63          & 0.66                   & 0.65                          \\ 
\hline
\multirow{3}{*}{\textbf{DL based}} & BERT\textsubscript{Base} & \multirow{3}{5cm}{batch\_size: 16, learning\_rate: 2e-05, weight\_decay: 0.01, optimizer: AdamW} & 0.83          & 0.80          & 0.82               & 0.78          & 0.82          & 0.80                   & 0.81                          \\
                                   & BERT\textsubscript{POS}  &                                                                                                & 0.82 & 0.76          & 0.79               & 0.71          & 0.83          & 0.77                   & 0.78                               \\
                                   & BERT\textsubscript{DEP} (CiRA)  &                                                                                                & \textbf{0.85}          & 0.81          & \textbf{0.83}      & 0.79          & \textbf{0.84} & \textbf{0.81}          & \textbf{0.82}              \\
\bottomrule
\end{tabular}}
\vspace{-.8cm}
\end{table}
\vspace{-.3cm} 
\section{Related Work}
\vspace{-.2cm}
As indicated in Section~\ref{terminology}, many disciplines have already dealt with causality. To the best of our knowledge, we are the first to focus on causality from the perspective of RE. In our previous paper~\cite{fischbach2020}, we motivated why the RE community should engage with causality, while in this paper we provide empirical evidence for the relevance of causality in requirement documents and an insight into its form and complexity. Detecting causality in natural language has been investigated by several studies. Multiple papers~\cite{khoo-etal-2000-extracting,wu05} use handcrafted patterns to identify causal sentences. These approaches are highly dependent on the manually created patterns and show weak performance. Recent papers apply neural networks and exploit, similarly to us, the Transfer Learning capability of BERT~\cite{kyriakakis2019transfer}. However, we see a number of problems with these papers regarding the realization of our described RE use cases: First, neither the code nor a demo is published, making it difficult to reproduce the results and testing the performance on RE data. Second, they train and evaluate their approaches on strongly unbalanced data sets with causal to non-causal ratios of 1:2 and 1:3, but only report the macro-Recall and macro-Precision values and not the metrics per class. Thus, it is not clear whether the classifier has a bias towards the majority class or not.
\vspace{-.4cm} 
\section{Conclusion and Next Steps}
\vspace{-.2cm}
System behavior is often specified by causal relations in requirements. Extracting this causal knowledge supports automatic test case derivation and reasoning about requirement dependencies~\cite{fischbach2020}. However, existing methods fail to extract causality with reasonable performance~\cite{Asghar16}. Therefore, we argue for the need of a novel method for causality extraction. We understand causality extraction as a two-step problem: First, we need to detect if requirements have causal properties. Second, we need to comprehend and extract their causal relations. At present, however, we lack knowledge about the form and complexity of causality in requirements, which is needed to develop suitable approaches for these two problems. In this paper, we address this research gap and contribute: (1) an exploratory case study with 14,983 sentences from 53 requirements documents originating from 18 different domains. We found that causality is a widely used linguistic pattern to describe system functionalities and that it mainly occurs in explicit, marked form. (2) CiRA as an approach for the automatic detection of causality in requirements documents. This constitutes a first step towards causality extraction from NL requirements. We empirically evaluate our approach and achieve a macro-F\textsubscript{1} score of 82~\% on real word data. (3) we disclose our code, tool and annotated data set to facilitate replication. 

Two further research directions exist: First, extending the case study and analyzing the sentences from the requirements documents in a more granular way by categorizing them e.g. in functional and non-functional requirements. This would expand our current insight into causality in requirements documents in general by an insight into causality in specific requirement categories. Second, we are enhancing our previous approaches~\cite{fischbach20,frattini2020} to address the second sub-problem: the actual extraction of causal relations. 
\vspace{-.3cm} 
\section*{Acknowledgements}
\vspace{-.3cm}
We would like to acknowledge that this work was supported by the KKS foundation through the S.E.R.T. Research Profile project at Blekinge Institute of Technology. Further, we thank Yannick Debes for his valuable feedback.
\vspace{-.3cm} 
\bibliographystyle{splncs04}
\bibliography{references}
\end{document}